\documentclass{article}
\usepackage{spconf,amsmath,graphicx,subfigure}
\usepackage{hyperref, booktabs}

\ninept
\title{Multi-speaker Multi-lingual VQTTS System for LIMMITS 2023 Challenge}
\name{Chenpeng Du$^\dagger$\thanks{$^\dagger$Equal contribution.}, Yiwei Guo$^\dagger$, Feiyu Shen$^\dagger$, Kai Yu$^\ddagger$\thanks{$^\ddagger$Corresponding author.}}
\address{MoE Key Lab of Artificial Intelligence, AI Institute\\X-LANCE Lab, Department of Computer Science and Engineering\\Shanghai Jiao Tong University, Shanghai, China\\ \small\texttt{\{duchenpeng, cantabile\_kwok, francis\_sfy, kai.yu\}@sjtu.edu.cn}}
\begin{document}
\ninept
\maketitle
\begin{abstract}
In this paper, we describe the systems developed by the SJTU X-LANCE team for LIMMITS 2023 Challenge, and we mainly focus on the winning system on naturalness for track 1. The aim of this challenge is to build a multi-speaker multi-lingual text-to-speech (TTS) system for Marathi, Hindi and Telugu. Each of the languages has a male and a female speaker in the given dataset. In track 1, only 5 hours data from each speaker can be selected to train the TTS model. Our system is based on the recently proposed VQTTS that utilizes VQ acoustic feature rather than mel-spectrogram. We introduce additional speaker embeddings and language embeddings to VQTTS for controlling the speaker and language information. In the cross-lingual evaluations where we need to synthesize speech in a cross-lingual speaker's voice, we provide a native speaker's embedding to the acoustic model and the target speaker's embedding to the vocoder. In the subjective MOS listening test on naturalness, our system achieves 4.77 which ranks first.
\end{abstract}
%

% \vspace{-.1in}

\section{Introduction}
% \vspace{-.1in}

\label{sec:intro}

The LIMMITS 2023 Challenge is organized as part of ICASSP 2023 which aims at the development of a lightweight multi-speaker multi-lingual Indic text-to-speech (TTS) model using datasets in Marathi, Hindi and Telugu. Each of the 3 languages has a male and a female speaker in the given dataset, so there are a total of 6 speakers. Each speaker has 40 hours speech data and the corresponding transcript.
In track 1, participants may only use at most 5 hours data from each speaker for training the TTS model. Track 2 focuses on the model size. It specifies mel-spectrogram as the acoustic feature and restricts the number of parameters in text-to-mel model to 5 million. In track 3, both the constraints on the amount of training data and the number of model parameters are imposed.

 For track 2 and track 3 where the model size is limited, we use GradTTS \cite{gradtts} that leverages diffusion model for mel-spectrogram prediction. The denoising process has a large number of infinitesimal steps but shares a same U-net architecture, so it can achieve high voice quality with limited number of parameters. 
 We reduce the parameter size from original 15 million to less than 5 million by using less layers in text encoder and lower channels in the U-net.
 EMA is applied on model parameters. 
 We also add speaker embedding and language embedding to the encoder output for controlling the speaker and language information. In the objective evaluations on naturalness, our system ranks first in track 3 with a MOS score of 4.44, and third in track 2 with MOS score 4.40.

In track 1, we choose VQTTS \cite{VQTTS}-based system that utilizes VQ acoustic feature rather than mel-spectrogram for robust and high-fidelity TTS training.
In the subjective MOS listening test on naturalness, our system achieves 4.77 which also ranks first. 
Note that this MOS score is already very close to human-level naturalness.
Next, we describe the system in detail in the rest of this paper.

\begin{figure*}[t]
  \centering
  \subfigure[txt2vec]{
    \includegraphics[width=0.4\linewidth]{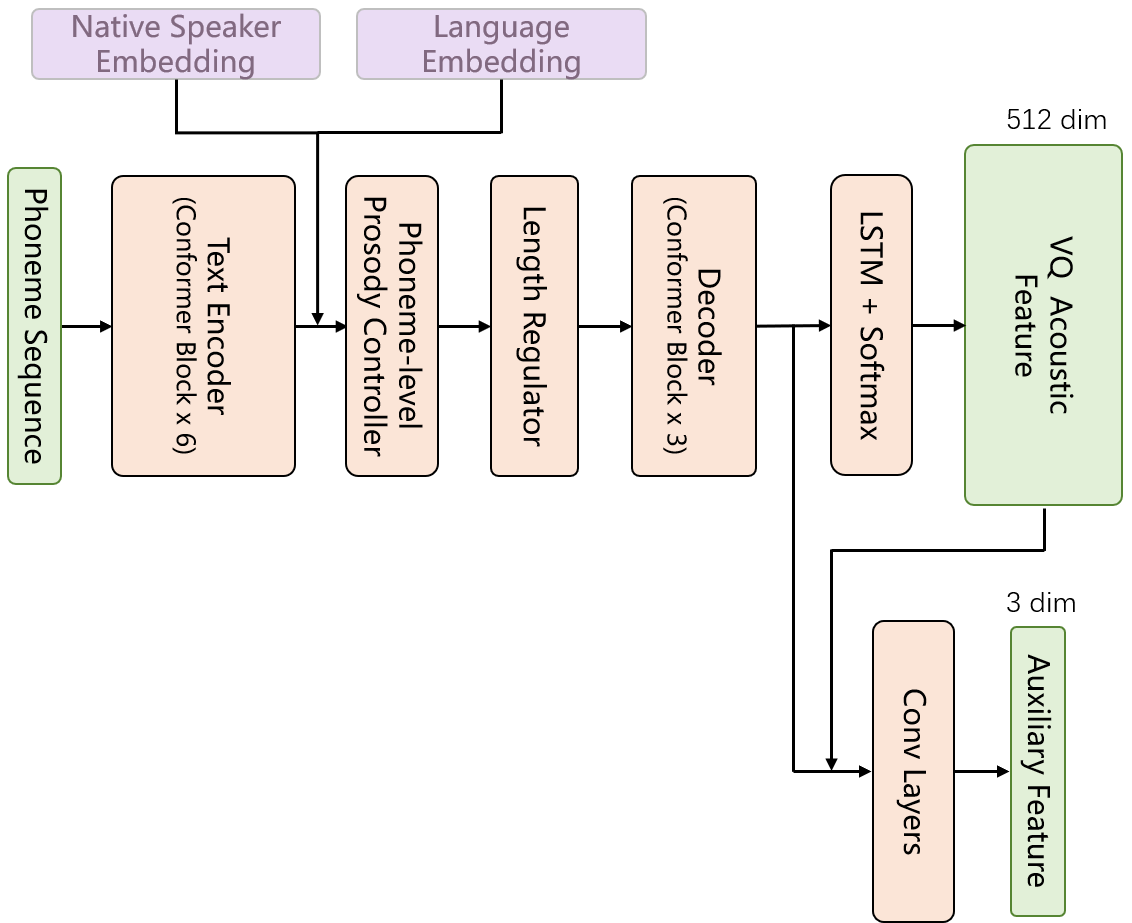}
    \label{txt2vec}
  }
  \subfigure[vec2wav]{
    \includegraphics[width=0.37\linewidth]{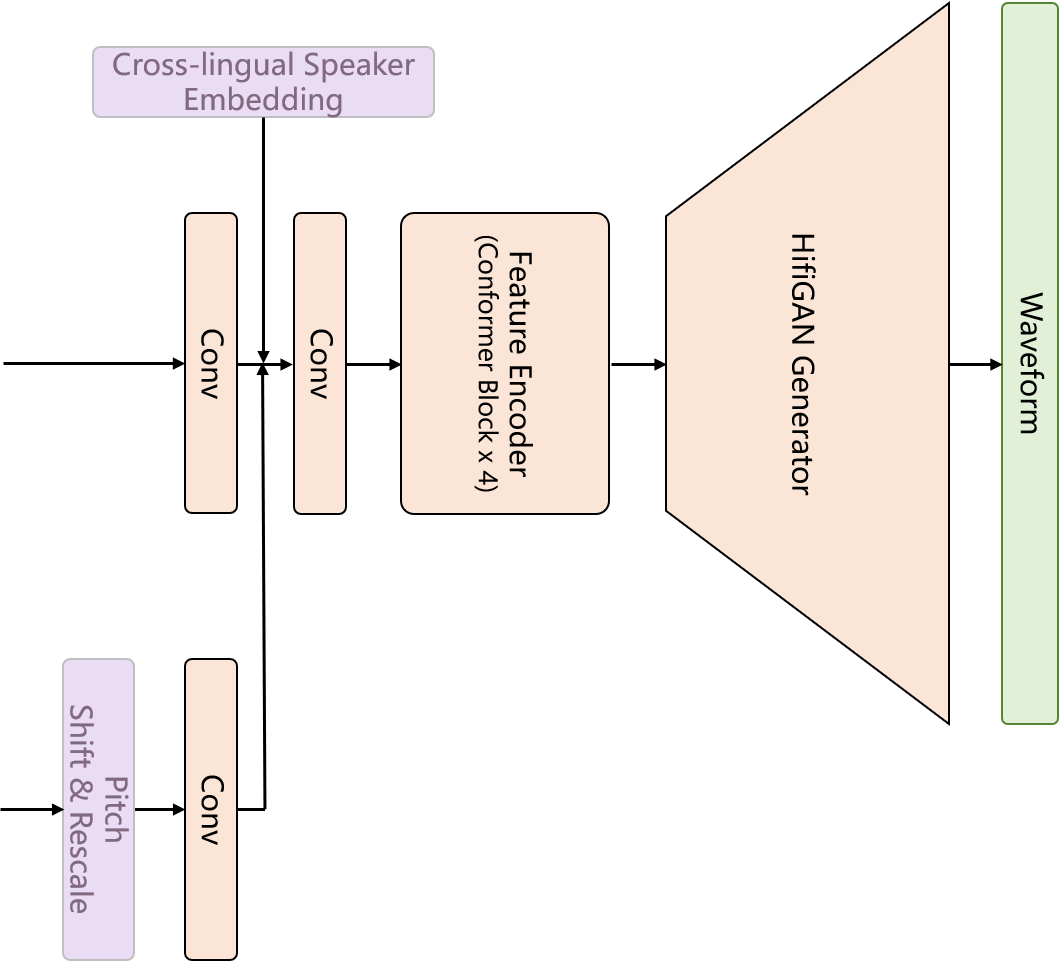}
    \label{vec2wav}
  }
  \caption{Cross-lingual speech synthesis in multi-speaker multi-lingual VQTTS.}
  \label{fig:VQTTS}
  \vspace{-0.3cm}
\end{figure*}

% \vspace{-.1in}
\section{System Description}
\label{sec:system}
% \vspace{-.1in}

\subsection{Data selection}

In track 1, only 5 hours data from each speaker can be used to train the TTS model, so we need to select the training data from the given total 40 hours data per speaker. First, we recognize the all speech data with the ASR model provided by the challenge organizers and compute the CER for each utterance between the transcript and the recognition result. We also train an alignment search model \cite{radtts} and computes the likelihood of the alignment for each utterance. Accordingly, we select 10 hours data that has the lowest CER and the highest likelihood of its alignment. Then we calculate the focus rate of the alignment for the selected 10 hours data and further select the best 5 hours data for TTS training.

% \vspace{-0.2in}
\subsection{Text preprocessing}
% \vspace{-.1in}

The three Indian languages can all be regarded as ideal phonemic orthography, so it suffices to use raw characters as TTS inputs. We find explicit schwa deletion is also not necessary in preliminary experiments. Note that we substitute digits with their pronunciations which are surrounded by braces in the provided transcripts.

Additional to raw characters, we also introduce an additional \texttt{<sil>} token that indicates a short silence. In order to detect the position of \texttt{<sil>} automatically in the training data, we train an HMM-GMM ASR model with Kaldi toolkit \cite{kaldi} and compute the alignment with token-passing algorithm on the WFST decoding graph. There is an optional silence between all the two consecutive words, so the algorithm automatically decides whether to go through the silence nodes for maximizing the likelihood. Given the training data with auto-detected silence position, we train a Transformer-based model to predict whether there are \texttt{<sil>} tokens between all the two consecutive words from raw text. In inference, we insert the \texttt{<sil>} tokens at the predicted positions into the character sequence and send the result to the TTS model.

From the Kaldi alignment, we also obtain the durations of all input tokens, which is used as the target for training the duration predictor of the TTS model.

\subsection{Multi-speaker multi-lingual VQTTS}

Our TTS model is based on the recently proposed VQTTS \cite{VQTTS}, consisting of an acoustic model txt2vec and a vocoder vec2wav. It uses VQ acoustic feature extracted by vq-wav2vec \cite{vqw2v} rather than mel-spectrogram as the acoustic feature. In this way, our acoustic model txt2vec basically becomes a classification model rather than a traditional regression model. 
  Instead of predicting the complicated mel-spectrogram with high correlation along both time and frequency axes, txt2vec only needs to consider the correlation along time axis in feature prediction, which narrows the gap between GT and predicted acoustic feature dramatically. Besides the VQ feature, VQTTS also introduces three dimensional auxiliary features including pitch, energy and POV.
  The vocoder vec2wav uses an additional feature encoder before HifiGAN generator for smoothing the discontinuous quantized feature. 

  To enable the model to control the speaker identity and language, we add speaker embeddings and language embeddings to the text encoder output. The speaker embeddings are also provided to the vocoder to control the generated voice. Our speaker embeddings are extracted by a pretrained x-vector extractor\footnote{https://huggingface.co/speechbrain/spkrec-ecapa-voxceleb} for all utterances. 
  Then we average the speaker embeddings over all utterances within each speaker and take the result as the input to the TTS model. The language embeddings are stored in a trainable lookup table and are jointly trained with the TTS model.

\vspace{-.1in}
\subsection{Cross-lingual speech synthesis}
There are evaluations on both mono-lingual and cross-lingual speech synthesis in this challenge. For mono-lingual speech synthesis, we just provide the specified speaker embedding and language embedding to the VQTTS. As for cross-lingual speech synthesis, we notice that giving unmatched speaker embedding and language embedding to the acoustic model txt2vec sometimes generates unstable results and mispronunciation in our preliminary experiments. Therefore, we provide a native speaker's embedding to txt2vec for generating the proper articulation first. Then we provide the specified cross-lingual speaker's embedding to the vec2wav for generating the voice of the correct speaker. 
We find providing different speaker embeddings as such helps improve the pronunciation to a large extent, indicating the potential of VQTTS to automatically disentangle acoustic and linguistic information.
Note that the predicted pitch of the native speaker in the auxiliary feature is also shifted and rescaled to the specified speaker and then sent to the vocoder. The pipeline is demonstrated in Figure \ref{fig:VQTTS}.

\vspace{-0.1in}
\section{Evaluation Result}

In the evaluation, we need to synthesize speech given specified content, speaker and language, including both mono-lingual and cross-lingual combinations of speaker and language. The MOS scores in the subjective listening tests are shown in Table \ref{tab:result}.

\vspace{-0.2in}

\begin{table}[ht]
\caption{Mean MOS scores of multi-speaker multi-lingual VQTTS.}
\label{tab:result}
\centering
\begin{tabular}{c|ccc}
\hline
 & All & Mono-lingual  & Cross-lingual \\ \hline
Nautralness         &  4.77   & 4.80   &  4.74 \\
Speaker Similarity  & 3.86    & 4.25   &  3.45 \\ \hline 

\end{tabular}
\end{table}

 Our proposed VQTTS system achieves the best naturalness among all the systems in track 1. We notice that the speaker similarity of cross-lingual synthesis is still worse than that of mono-lingual synthesis. We leave this issue to be addressed in the future work.

\vspace{-.1in}

\bibliographystyle{IEEEtran}
\bibliography{strings,refs}

\end{document}